\documentclass[12pt,letterpaper]{article}
\pdfoutput=1
\usepackage{graphicx,array}
\usepackage{color}
\usepackage{latexsym}
\usepackage{amsthm}
\usepackage{amsmath}
\usepackage{enumitem}
\usepackage{amssymb}
\usepackage{hyperref}
\usepackage[hang,flushmargin]{footmisc}

\def\simgt{\mathrel{\lower2.5pt\vbox{\lineskip=0pt\baselineskip=0pt
           \hbox{$>$}\hbox{$\sim$}}}}
\def\simlt{\mathrel{\lower2.5pt\vbox{\lineskip=0pt\baselineskip=0pt
           \hbox{$<$}\hbox{$\sim$}}}}

\newcommand{\be}{\begin{equation}}
\newcommand{\ee}{\end{equation}}
\newcommand{\bea}{\begin{eqnarray}}
\newcommand{\eea}{\end{eqnarray}}
\newcommand{\Eq}[1]{Eq.~(\ref{#1})}

\newcommand{\Sec}[1]{Sec.~\ref{#1}}

\newcommand{\Fig}[1]{Fig.~(\ref{#1})}

\newcommand{\vev}[1]{\langle #1 \rangle}

\newcommand{\mPl}{m_{\rm Pl}}

\newcommand{\sslash}[1]{\ensuremath\raisebox{-0.00cm}{{\small\slash}}\hspace{-0.21cm}#1\/}

\definecolor{nicered}{rgb}{0.7,0.1,0.1}
\definecolor{nicegreen}{rgb}{0.1,0.5,0.1}
\hypersetup{colorlinks,citecolor= black,linkcolor= black}

\begin{document}
\hfill

\vspace{4cm}

\begin{center}
{\LARGE\bf
Limits on New Physics from Black Holes
}\\
\bigskip\vspace{1cm}{
{\large Clifford Cheung and Stefan Leichenauer}
} \\[7mm]
 {\it California Institute of Technology, Pasadena, CA 91125}
 \end{center}
\bigskip
\centerline{\large\bf Abstract}

\begin{quote} \small

Black holes emit high energy particles which induce a finite density potential for any scalar field $\phi$ coupling to the emitted quanta.  Due to energetic considerations, $\phi$ evolves locally to minimize the effective masses of the outgoing states.  In theories where $\phi$ resides at a metastable minimum, this effect can drive $\phi$ over its potential barrier and classically catalyze the decay of the vacuum.  Because this is not a tunneling process, the decay rate is not exponentially suppressed and a single black hole in our past light cone may be sufficient to activate the decay.  Moreover, decaying black holes radiate at ever higher temperatures, so they eventually probe the full spectrum of particles coupling to $\phi$.  We present a detailed analysis of vacuum decay catalyzed by a single particle, as well as by a black hole.  The former is possible provided large couplings or a weak potential barrier.  In contrast, the latter occurs much more easily and places new stringent limits on theories with hierarchical spectra.  Finally, we comment on how these constraints apply to the standard model and its extensions, {\it e.g.}~metastable supersymmetry breaking.
\end{quote}

\newpage

\tableofcontents

\newpage

\section{Introduction }


Black holes are a naturally occurring source of high energy particles.  
During evaporation, a black hole emits a continuous flux of Hawking radiation which in steady state forms a halo of free-streaming particles.  If this distribution is sufficiently dense it will influence the dynamics of a scalar field coupled to the outgoing states.   Away from the horizon the emitted quanta are out of thermal equilibrium, so they induce a finite density, zero temperature potential for the scalar.  While the precise form of this potential depends on the microscopic dynamics, the scalar field always moves to minimize the effective masses of the emitted quanta.  This phenomenon is reminiscent of the bag mechanism discussed in \cite{Khlebnikov:1986ky,Dimopoulos:1990at,Anderson:1990kb}, only here it is viable in weakly coupled theories, provided there is a sufficiently large flux of particles created by the black hole.

If the scalar field resides in a metastable vacuum, then it is possible for the finite density potential to overcome the potential barrier and catalyze the decay of the vacuum.  In this paper we present a detailed analysis of this mechanism and its implications for new physics.   While this idea has been studied in the context of weakly coupled string moduli \cite{Green:2006nv}, we believe it has much broader applications to physics beyond the standard model (SM).   As we will see, black hole catalyzed vacuum decay can place powerful new limits on theories which have in the past been deemed safe by conventional stability bounds.  This is true because:

\begin{itemize}[leftmargin=.5cm,rightmargin=.5cm] 

\item {\bf Vacuum decay is classical.}  
This mechanism utilizes the classical activation of a scalar field over its potential barrier.  Unlike for quantum tunneling, the associated decay rate is not exponentially suppressed.  Consequently, a single black hole in our past light cone can be sufficient to destabilize the vacuum.  This constrains many models which in empty space are metastable with lifetimes longer than the age of the universe.

\item {\bf Black holes get very hot.} As a black hole decays, its temperature scans adiabatically up to the Planck scale.  There is no kinematic limit for high energy particle production because every mass threshold is accessible to a sufficiently hot black hole.  Contrast this with limits from thermally assisted vacuum decay in the early universe, which depend on the initial reheating temperature.


\end{itemize}
Catalyzed vacuum decay remains a relatively unexplored topic.  The authors of \cite{Affleck:1979px,Voloshin:1993dk} considered catalysis by individual particles, concluding that the decay rate is only modestly enhanced.  Meanwhile, \cite{Berezin:1990qs,Arnold:1989cq} analyzed  vacuum decay in the presence of a black hole,  incorporating the effects of the metric but neglecting the effects of the emitted Hawking radiation.  Critically, past examples have emphasized quantum mechanical tunneling rather than the mechanism of classical activation discussed in this paper and \cite{Green:2006nv}.

The outline of this paper is as follows.  In \Sec{sec:BH} we compute the phase space density of Hawking radiation emitted by a black hole.  We then derive a simple formula for the finite density potential.  Afterwards, in \Sec{sec:decay} we determine the conditions under which vacuum decay is catalyzed by a point particle, and by the halo of quanta surrounding a black hole.  Finally, in \Sec{sec:pheno} we discuss the implications of catalysis for a concrete model, as well as for the SM and its extensions.  We summarize our conclusions and discuss prospects for future work in \Sec{sec:conclusions}.

\section{Finite Density Potential}

\label{sec:BH}

In this section we compute the phase space density of Hawking radiation far from the black hole horizon. We also demonstrate that these particles are effectively free-streaming and out of thermal equilibrium.  Finally, we derive a simple formula for the finite density potential, using independent methods from classical and quantum mechanics.

\subsection{Hawking Radiation Distribution}

Black holes emit Hawking radiation at a rate of  \cite{Hawking:1974rv,Hawking:1974sw,Unruh:1976fm,Page:1976df}
\bea
\Gamma = \int d^3 k\; \xi(k) \qquad , \qquad
\xi(k)
=\frac{1}{(2\pi)^3} \; \frac{\sigma  v}{e^{w/T}\mp 1},
\label{eq:Gamma}
\eea
where $T$ is the temperature of the black hole and $\sigma$ is the absorption cross-section of the black hole with respect to the emitted  particle.  As is well-known, the spectrum of Hawking radiation is only approximately thermal due to grey body factors encoding the dependence on particle species in the absorption cross-section \cite{Unruh:1976fm,Page:1976df}.  For example, the emission rate is lower for particles of higher spin.

We characterize the outgoing flux of particles with a phase space density, $n(k,x)$, describing an ensemble of classical particles with well-defined momentum and position.   We can crudely estimate $n(k, x)$ by equating the number of particles in a given frequency band and infinitesimal volume region with the flux of Hawking radiation emitted at the horizon in an infinitesimal time interval, 
\bea
dN = n(k, x)   \; d^3 k\; d^3 x = \Gamma \; dt = \xi(k) \; d^3k \; dt,
\eea
Here we ignore metric effects because we are interested in particles far from the horizon.    Of course, gravitational redshift should still be included in the calculation of the initial black  hole emission spectrum.

If we assume that the emitted particles free stream in the radial direction, then the resulting phase space density is spherically symmetric and 
\bea
 n(k, r)  &=&  \frac{\xi(k)}{4\pi r^2}\frac{\omega}{k},
 \label{eq:number}
\eea
where we have assumed that the distribution is in steady state so that we can set  $dr/dt$ equal to the velocity of the emitted relativistic massive particle, $v= k/\omega$.  Naively $n(k,r)$ diverges at low velocities, but this spurious singularity is cancelled when computing physical quantities like the spectrally integrated number density, $n(x) = \int d^3k\; n(k,x)$.

\Eq{eq:number} can be derived in a less heuristic way.  The phase space distribution is equal to a spectrally weighted sum of delta functions, each localized to the position of an emitted particle.  Consider a time period $t \in [0,\tau ]$, where $\tau$ is a fiducial time interval over which the emission spectrum can be treated as constant.  We find that
\bea
n(k, r) &=& \xi(k) \int_0^{\tau} dt\; \delta^3(r -  v  t)\nonumber\\
&=&    \frac{\xi(k)}{4\pi r^2}  \frac{\omega}{k} \times\theta\left(\frac{k}{\omega} -\frac{ r}{\tau }\right),
\eea
which for $\tau  \rightarrow \infty$ asymptotes to the more crudely derived estimate in \Eq{eq:number}.   Note that the large $\tau$  approximation is justified  because black holes decay on extremely long time scales relative to their characteristic size.  Concretely, the black hole lifetime is $\tau_{\rm BH} = 5120\pi G^2 M_{\rm BH}^3$, which in units of the Schwarzschild radius is
\bea
\frac{\tau_{\rm BH}}{R_{\rm BH}} = 640 \pi G^{-1} R^2_{\rm BH} \gg 1.
\label{eq:BHdecay}
\eea
We will return to the issue of the black hole lifetime later on in \Sec{sec:decay}.

Our derivation of the phase space density has made use of several critical assumptions: 1) the mean field approximation is valid, 2) the emitted particles are free-streaming, and 3) they are sufficiently long-lived that the number density reaches a steady state.  Let us scrutinize each of these assumptions in turn.

First of all, the mean field description is only justified if $n(k,x)$ describes a large number of particles.  Concretely, we require that within a region of size $r\gg R_{\rm BH}$, the total number of particles is
\bea
N &=& \int d^3k \; d^3 x \; n(k,x) \nonumber\\ &=& r \Gamma \sim r/R_{\rm BH}(4\pi)^4 \gg 1 ,
 \label{eq:denseenough}
\eea
where in the last line we have inserted $\Gamma \sim 1/R_{\rm BH}(4\pi)^4$, which is a parametric estimate of \Eq{eq:Gamma} taking into account $4\pi$ factors.  Thus, as long as we restrict to distances very far from the event horizon and only consider dynamics on length scales much larger than $R_{\rm BH}$, the mean field approximation applies.

Second, we must ascertain whether the emitted particles can be treated as free particles after they are emitted.  The scattering rate amongst relativistic emitted particles is 
\bea
\Gamma_{\rm scatt} &=& n(x) \sigma_{\rm scatt}\nonumber \\
&=&  \Gamma \times \frac{ \sigma_{\rm scatt} }{4\pi r^2} ,
\eea
where $n(x)$ is the spectrally integrated phase space density.  The scattering rate scales with the number density, which decreases at large radii.  Assuming a perturbative cross-section of order $\sigma_{\rm scatt} \sim  4\pi \alpha^2 /T^2$, we find that
\bea
\Gamma_{\rm scatt} &=& \Gamma \times (4\pi \alpha)^2\left(\frac{R_{\rm BH}}{r}\right)^2 .
\eea
The number of times an emitted particle scatters before it escapes to infinity is
\bea
N_{\rm scatt} &=& \int_R^\infty dr \; \Gamma_{\rm scatt} \nonumber \\
&=&  \Gamma R_{\rm BH}  \times (4\pi \alpha)^2 \sim (\alpha/4\pi)^2  < 1 .
\label{eq:nothermal}
\eea
Hence, the flux of Hawking radiation does not thermalize.  These estimates are consistent with \cite{MacGibbon:2007yq}, which reached a similar conclusion.

Third, let us consider the lifetime of the emitted particles.  For simplicity, we assume a perturbative decay rate,
\bea
\Gamma_{\rm dec} &\sim&  \alpha M  ,
\eea
where $M$ is the mass of the Hawking radiation.  Crucially, the associated decay length is enhanced by a substantial boost factor, $\gamma$, because the particles are relativistic.  For the phase space density of emitted particles to reach a steady state, the decay length is bounded by  $R_{\rm dec} = \Gamma_{\rm dec}^{-1} \gamma \gg \Gamma^{-1}$ so that the rate of particle depletion is overcome by the rate of production by Hawking radiation.  This implies the parametric condition, $\gamma \gtrsim \sqrt{\alpha}$, which is easily satisfied for relativistic particles.

\subsection{Classical Derivation}
\label{sec:Veff}

Consider a scalar field, $\phi$, which couples to the outgoing Hawking radiation, $\chi$.  We now present a classical derivation of the finite density potential for $\phi$ induced the ambient $\chi$ particles. 
To begin, consider a  microscopic Lagrangian density, ${\cal L}$, containing  arbitrary interactions between $\phi$ and $\chi$.  In a regime where $\phi$ is slowly varying relative to $\chi$ we can define a $\phi$-dependent, effective mass squared for $\chi$,
\bea
\mu^2[\phi] &=& -\frac{\partial^2 {\cal L}}{\partial \chi^2}\left(\frac{\partial^2 {\cal L}}{ \partial \dot{\chi}^2}\right)^{-1/2}.
\label{eq:Meff}
\eea
We ignore metric effects because we are interested physics far from the horizon, where gravitational redshift is negligible.\footnote{It is possible that a large correction to the effective potential in the near-horizon region could alter our conclusions, but since the near-horizon region is parametrically small compared to our region of interest we do not believe this will be the case.}  The first factor in \Eq{eq:Meff} denotes the $\phi$-dependent ``bare mass'' induced by non-derivative couplings such as $\phi \chi^2$.  The second factor denotes the $\phi$-dependent ``wavefunction renormalization'' induced by derivative couplings such as $\phi \partial \chi^2$.  While \Eq{eq:Meff} applies to a real scalar $\chi$, the generalization to complex or higher spin fields is obvious.  

Thus far we have only considered terms quadratic in $\chi$, but a general theory will also include couplings that mediate $\chi$ self-interactions and decays.  However, since the emitted $\chi$ particles are freely propagating in our regime of interest, $\chi$ particle number is effectively conserved and these interactions can be ignored.   We will return to the issue of $\chi$ decays later on.

Consider a regime in which $\phi$ is slowly varying relative to the momenta of the background $\chi$ particles.  The ensemble of $\chi$ particles is described by a worldline action, $S = \int dt\; L$, where
\bea
S &=&  - \sum_i \int ds_i \; \mu[\phi(t, x_i)] \nonumber\\
&=& - \sum_i \int dt \;  \mu[\phi(t, x_i)]\sqrt{1- \dot  x_{i} ^2},
\eea
where $i$ labels each $\chi$ particle and  $(t, x_i)$ its worldline trajectory.
The canonical momentum for each particle is $k_i = \partial L/\partial x_i = \mu[\phi(t, x_i)] \dot x_i / \sqrt{1- \dot x_i^2}$ and the associated Hamiltonian is
\bea
H &=& \sum_i   \sqrt{k_i^2 + \mu^2[\phi(t, x_i)]} \nonumber\\
&=& \int d^3 x \sum_i \sqrt{k_i^2 + \mu^2[\phi(t, x)]}\times  \delta^3(x- x_i),
\eea
which is the finite density analog of the Coleman-Weinberg potential.
We will only consider situations where the distance between $\chi$ particles is much less than the length scales in the $\phi$ potential. Then it is valid to take the continuum limit,
\bea
H&=&\int d^3 x \int d^3k \; \sqrt{k^2 + \mu^2[\phi(t, x)]} \times n(k,x) \nonumber \\
&=&\int d^3 x \; \mu^2 [\phi(t, x)] \times \frac{1}{2} \int d^3k  \;  n(k,x) \left(\frac{1}{k}+ {\cal O}(1/k^3)\right),
\label{eq:Heff}
\eea
where $n(k,x)$ is the phase space density and we have expanded around the relativistic limit.  
Plugging in our expression for $n(k,x)$ from \Eq{eq:number} into \Eq{eq:Heff}, we obtain a simple expression for the finite density potential induced by a black hole,
\bea
\Delta V_{\rm BH} &=&  \frac{f}{ r^2} \times \mu^2[\phi],
\label{eq:Veff}
\eea
which is valid in the limit of relativistic $\chi$ particles.  Here the dimensionless number $f$ encodes the effects of grey body factors,
\bea
f &=&  \frac{1}{8 \pi}  \int d^3k  \;  \frac{\xi(k)}{k} \nonumber \\
 &=&\frac{1}{16\pi^3 } \int dk \; \frac{k\sigma(k,T)}{e^{k/T}\mp 1} ,
\eea
and we have set $\omega=k$ throughout.  \Eq{eq:Veff} makes sense physically: $\phi$ is driven towards field values that minimize the effective masses of ambient $\chi$ particles.  However, this effect falls off at large distances due to the dissipation of the flux of Hawking radiation.

By dimensional analysis $f$ is independent of $T$, but it varies with the spin of the emitted particle.
Employing the approach of \cite{Harris:2003eg}, we have calculated $f$ from  black hole grey body factors determined by the absorption cross-section of a particle incident on the black hole.  Specifically, we solved numerically for the reflection and transmission coefficients of an incoming wave of a Klein-Gordon field propagating in a Schwarzschild background.  Our results closely match those of \cite{Harris:2003eg}, and are consistent with the approximate analytic expressions in \cite{Cvetic:1997ap}.   We find that
\bea
f
&\simeq & 10^{-4} \times \left\{
\begin{array}{ll}
3.9 \qquad &,\qquad  s=0 \\ \\
0.7 \qquad &,\qquad  s=1/2 \\ \\
0.1 \qquad &,\qquad  s=1 \\ 
\end{array} \right. 
\label{eq:fTexact}
\eea
As is well known, higher spin particles have a suppressed rate of emission.

\subsection{Quantum Derivation}

\Eq{eq:Veff} can also be derived quantum mechanically.  In the approximation that $\phi$ is a slowly varying background for $\chi$, the microscopic Hamiltonian density is
\bea
{\cal H} &=& \frac{1}{2} \dot \chi^2 + \frac{1}{2}\nabla \chi^2 + 
\frac{1}{2}  \mu^2[\phi] \chi^2 ,
\label{eq:energydensity}
\eea
where the kinetic terms have been canonically normalized.
Near a black hole, the exiting $\chi$ quanta are relativistic and their total energy is dominated by their $\phi$-independent kinetic energy.  Consequently, the leading order $\phi$ dependence arises from the effective mass term in \Eq{eq:energydensity}.  The finite density potential is then determined by the usual rules of quantum mechanical perturbation theory, so
\be
\Delta V_{\rm BH} =\frac{1}{2} \mu^2[\phi] \vev{\chi^2},
\label{eq:QM1}
\ee
where the expectation value is evaluated on the wavefunction characterizing the outgoing radiation. 

Classically, we know that the number density of outgoing particles reaches an approximate steady-state, and so the expectation value of \Eq{eq:QM1} must be constant in time. In the interaction picture, the wavefunction itself for free-streaming particles is also constant. There is a trivial sort of time-dependence coming from the continual emission of new particles from the black hole, but this only serves to replace the particles which are free-streaming away. Therefore we are justified in approximating the wavefunction in \Eq{eq:QM1} as time-independent. As we will see, the precise form of this wavefunction is unimportant, provided the true wavefunction describes classical, free-streaming particles far from the black hole. Note that we are explicitly neglecting any transient effects from the black hole formation, as well as slow changes occurring on the timescale of the black hole lifetime and all nontrivial metric effects (including interactions with the Newtonian gravitational potential). These should all be good approximations over the length, time, and energy scales we are interested in. In particular, though we are not computing the effective potential in the near-horizon region, we are assuming that its net effect on vacuum decay over much larger distances is negligible. 

Let us compute the expectation value of $\chi^2$ on a single particle state, $|\psi \rangle = \int d^3k\; \psi(k) |k\rangle$, using the normalizations $\langle k' | k \rangle = \delta^3(k-k')$ and $\int d^3k\; |\psi(k)|^2 = 1$.  A short calculation yields
\bea
\langle \psi |\chi(t,x)^2 |\psi\rangle &=& \int \frac{d^3k_1 \; d^3k_2}{(2\pi)^3}\; \frac{\psi^*(k_1)\psi(k_2)}{\sqrt{\omega_1 \omega_2}} e^{i(\omega_1 -\omega_2)t}e^{-i(k_1 -k_2)x},
\label{eq:chisq}
\eea
where $\omega_{1} \simeq k_{1}$ and $\omega_{2} \simeq k_{2}$ because the initial particle is relativistic.  Since we are interested in length and time scales much longer  the Compton wavelength of the relativistic particle, the momentum transfer is minute, so $k_1 \simeq k_2$.  Expanding the denominator of the integrand of \Eq{eq:chisq} as $\sqrt{\omega_1 \omega_2} = (\omega_1+\omega_2)/2 + {\cal O}(\omega_1 - \omega_2)$, we obtain
\bea
\langle \psi |\chi(t,x)^2 |\psi\rangle &\simeq& \int d^3k \; \frac{W(k,x)}{k} ,
\label{eq:chisq2}
\eea
where we have defined the quantum mechanical Wigner function,
\bea
W(k,x) &=& \int \frac{d^3k_1 \; d^3k_2}{(2\pi)^3}\; \psi^*(k_1)\psi(k_2) \delta^3(k-(k_1+k_2)/2)e^{i(\omega_1 -\omega_2)t}e^{-i(k_1 -k_2)x}.
\eea
As is well known, $W(k,x)$ is a Fourier transform of the density matrix which faithfully encodes all of the information of the wavefunction.   Expectation values are computed from the moments of $W(k,x)$ distribution.  Moreover, the Wigner function has the remarkable property that it asymptotes to the classical phase space distribution in the classical limit.  This limit applies in regions far from the event horizon, where the emitted quanta are free-streaming.  While \Eq{eq:chisq2} applies to the case of a single particle, we can straightforwardly accommodate the effects of $N$ independent particles by including a multiplicity factor.  We can then relate the Wigner function to the classical phase space density according to $W(k,x) = n(k,x)/N$, yielding the result of \Eq{eq:Heff}.  The quantum and classical derivations yield the same expression for the finite density potential.

\section{Catalyzed Vacuum Decay}

\label{sec:decay}

In this section we present a detailed analysis of catalyzed vacuum decay.  Parameterizing the metastable vacuum with a simple scalar field theory, we derive precise criteria for catalyzed decay by a point particle and by a black hole.  For the sake of generality, we express our results in terms the general finite density potential rather than any specific model.  

\subsection{Scalar Potential}

Consider a scalar field $\phi$ with a generic potential.  Without loss of generality, we choose the origin of the field space to be the location of the vacuum.  Expanding around the vacuum, the leading renormalizable potential is
\bea
V &=&  \frac{m^2 \phi^2}{2}  - \frac{a \phi^3}{3!} + \frac{\lambda\phi^4}{4!},
\label{eq:Vexplicit}
\eea
where $m^2>0$ so the origin is a local minimum, and $\lambda >0$ so the potential is bounded from below.  For later convenience, we go to ``hatted'' dimensionless variables,

\be
\begin{aligned}[c]
\hat x &= x m\\
\hat \phi &= \phi \sqrt{\lambda}/ m\\
\hat a &= a/\sqrt{\lambda} m \\
\end{aligned}
\qquad , \qquad
\begin{aligned}[c]
\hat V &=V \lambda /m^4  =
\frac{\hat \phi^2}{2}    - \frac{\hat a \hat \phi^3}{3!} + \frac{ \hat\phi^4}{4!}  
\end{aligned}
\label{eq:rescale}
\ee
The origin is metastable provided the cubic coupling is sufficiently large, 
\bea
\hat a   > \sqrt{3},
\label{eq:metastable}
\eea
in which case the  potential has local minima at
\bea
\hat \phi_{\rm false} &=& 0 \nonumber\\
\hat \phi_{\rm true} &=& 
 \frac{3}{2} \left(
   \sqrt{\hat a^2-8/3}+ \hat a\right).
   \label{eq:vac}
\eea
Since the minima are separated by a potential barrier, vacuum decay is mediated by quantum mechanical tunneling.  Occasionally, rare quantum fluctuations will nucleate a bubble of true vacuum.  Driven by the energy differential between minima, the bubble walls quickly accelerate  to near the speed of light and convert the entire universe to true vacuum.    As is well-known, the  decay rate is exponentially suppressed by the Euclidean action evaluated at the saddle point associated with the vacuum decay process \cite{Coleman:1977py,Callan:1977pt}.  
When \Eq{eq:metastable} approaches saturation, $\hat a \rightarrow \sqrt{3}$, the minima are approximately degenerate and the nucleated bubble is thin wall.  In the opposite regime, $\hat a \gg \sqrt{3}$, the potential barrier is weak and the nucleated bubble is thick wall. In either case, quantum mechanical vacuum decay is  exponentially slow because it requires tunneling from the metastable vacuum into a coherent field configuration.

%

\begin{figure}[t]\hspace*{-.5cm}
\centerline{\includegraphics[width=0.7\columnwidth]{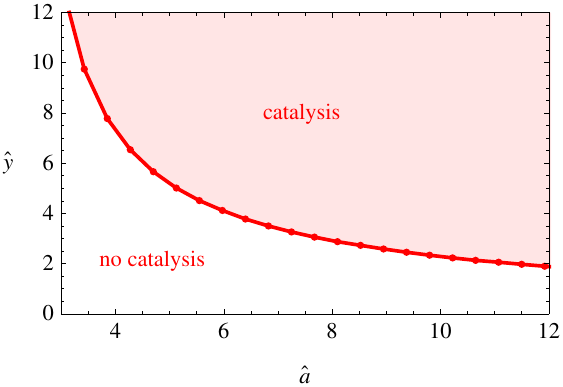}}
\caption{Phase diagram indicating when a single point particle classically catalyzes the decay of the vacuum.  Destabilization requires a very large coupling ($\hat y \gg 1$) or a very weak potential barrier ($\hat a \gg \sqrt{3}$).
}\label{fig:ycrit}
\end{figure}

\subsection{Point Particle Instability}

\label{sec:point}

In empty space, vacuum decay is exponentially slow.  Can it be accelerated in the presence of matter?  For concreteness, we introduce interactions between $\phi$ and a Dirac fermion 
$\chi$,
\bea
-{\cal L }_{\rm int} &=& (M-y\phi) \bar \chi \chi.
\label{eq:Lint}
\eea
It is illuminating to understand the most extreme possibility: vacuum decay catalyzed by a single $\chi$ particle.  At first glance this sounds impossible.  On the other hand, if $\chi$ is much heavier than the characteristic mass scale of the $\phi$ potential, then this prospect becomes less outlandish.  
When $M \gg m$, a single $\chi$ particle can be treated as a point particle source for $\phi$,
\bea
\Delta V_{\rm PP} &=& -y \phi \delta^3(x).
\eea
At quadratic order in the Lagrangian, this induces a Yukawa potential, 
\bea
\phi &=& \frac{y e^{-mr}}{4\pi r}.
\eea  
The $1/r$ divergence is regulated near $r \sim M^{-1}$, the Compton wavelength of $\chi$.  At small radii, $\phi$ can be much greater than than $m$, signifying a coherent state comprised of a large number of $\phi$ quanta.   This is precisely analogous to the huge number of photon quanta that comprise the classical electric field around an electron.   The Yukawa profile for $\phi$ can easily attain field values at or beyond the true vacuum.  In principle, this can drive $\phi$ over its potential barrier and catalyze the decay of the vacuum.  There are important subtleties, however.  First of all, the source term must overcome the gradient energy cost of nucleating a bubble of true vacuum.  Second, the Yukawa potential screens at $r\gtrsim m^{-1}$, so it is unclear that $\phi$ has support at sufficiently large radii to catalyze vacuum decay.  Naively, these effects may be compensated for by larger values of $y$, but the viability of catalyzed vacuum decay remains a detailed question.

A definitive answer requires an analysis of the $\phi$ equations of motion.  In particular,  vacuum decay is catalyzed if the equations of motion do not admit a
stable solution that interpolates to the metastable vacuum far from the black hole.  This proposition is equivalent to saying that every solution which connects to the false vacuum is unstable.    Because the system is rotationally symmetric and the metastable vacuum carries zero angular momentum, we restrict to radially symmetric solutions.  For our boundary conditions we impose $\phi( r \rightarrow \infty)=0$ so that $\phi$ asymptotes to the metastable vacuum at infinity, and $ \phi(r\rightarrow 0) =y/4\pi r$ so that $\phi$ interpolates correctly onto the Yukawa potential at short distances.   In dimensionless variables, the equation of motion is
\bea
 \hat\square \hat \phi + \hat \phi - \frac{\hat a \hat  \phi^2}{2}  + \frac{\hat\phi^3 }{3!}- \hat y  \delta^3(\hat x)  &=&0,
 \label{eq:eompoint}
\eea
where $\hat y = y\sqrt{\lambda}$.   While \Eq{eq:eompoint} is not analytically solvable, it is straightforwardly integrated via numerical methods\footnote{ \Eq{eq:eompoint} can be numerically solved by substituting $\phi = \Delta \phi +y/4\pi r$  to eliminate the delta function term.  }.  Doing so over a range values of the two model parameters, $\hat a$ and $\hat y$, we have determined when the system permits a radially symmetric ground state that asymptotes to the metastable vacuum at infinity.  Our results are presented in \Fig{fig:ycrit}, which is a phase diagram depicting the value of $\hat y$ above which the decay is destabilized.  \Fig{fig:ycrit} implies that point particle catalyzed vacuum decay is difficult.  Near the thin wall limit, the critical coupling is non-perturbative, so our perturbative analysis cannot be trusted.   While catalysis may be viable in the very thick wall limit, it requires very large values of the coupling.  

Obviously, catalyzed vacuum decay is more efficient in the presence of multiple $\chi$ particles.   Such a situation can arise in the early universe if the reheating temperature is greater than the mass of $\chi$.  However, the associated limits depend sensitively on the reheating temperature after inflation.  On the other hand, a black hole is a source of $\chi$ quanta whose temperature is essentially independent of the cosmological history of the universe.

\subsection{Black Hole Instability}

\label{sec:BHdecay}

In this section we determine when vacuum decay is catalyzed by a black hole.  Expanding $\Delta V_{\rm BH}$ around the metastable vacuum, we find
\bea
\Delta V_{\rm BH} &=&  \frac{f}{r^2}  \times \left(  \mu^2 +  \mu^{2\, \prime} \phi + \frac{1}{2}\mu^{2\, \prime\prime } \phi^2  + \ldots \right),
\label{eq:Vexpand}
\eea
introducing a shorthand notation where $\mu^2$, $\mu^{2\, \prime}$, and $\mu^{2\, \prime \prime}$ denote $\mu^2[\phi]$ and its derivatives evaluated at $\phi=0$.
Note that $\mu$ is equal to the mass of $\chi$ in the metastable vacuum.  Neglecting $\phi$-independent terms, there are two possibilities for a leading instability:
\begin{itemize}[leftmargin=.5cm,rightmargin=.5cm]

\item {\bf Tadpole ($\Delta V_{\rm BH} \sim \phi$).}  For generic couplings, $\mu^{2\,\prime} \neq 0$ and the leading potential term is a tadpole.  The basin of attraction of $\Delta V_{\rm BH}$ is misaligned from the metastable vacuum.  For the appropriate sign of $\mu^{2\,\prime}$, the field is driven towards the potential barrier.

\item {\bf Tachyon ($\Delta V_{\rm BH} \sim \phi^2$).}  For certain theories with additional symmetry, $\mu^{2\, \prime} = 0$ and the leading potential term is quadratic.  If $\mu^{2\, \prime\prime}<0$, then $\Delta V_{\rm BH}$ induces a local tachyon.

\end{itemize}
In \Sec{sec:tadpole} and \Sec{sec:tachyon} we analyze each of these possibilities in turn.   
As we will see, catalyzed vacuum decay is viable for the case of tadpoles, but not tachyons.

\subsubsection{Tadpole Instability}
\label{sec:tadpole}



Consider a scenario where the leading potential term of the finite density potential is a tadpole.  This is generic, absent special symmetries restricting the couplings of the theory.
Without loss of generality, this scenario is described by the potential in \Eq{eq:Vexplicit} together with the induced tadpole from the finite density potential.  In dimensionless units, the equations of motion are
\bea
 \hat\square \hat \phi + \hat \phi - \frac{\hat a \hat  \phi^2}{2}  + \frac{\hat\phi^3 }{3!}- \frac{\hat b}{\hat r^2}  &=&0,
 \label{eq:eom}
\eea
where  $\hat b$ parameterizes the effect of ambient $\chi$ particles on the metastable vacuum,
\bea
\hat b &=& -\frac{f \sqrt{\lambda} \mu^{2\, \prime}}{m}.
\label{eq:bdef}
\eea
Catalysis depends sensitively on the sign of $\mu^{2\, \prime}$ because this quantity  controls how the effective mass of $\chi$ varies with $\phi$.  In particular,  $\hat b > 0$ drives $\phi$ positive and towards the true vacuum, while vice versa for $\hat b <0$.


When $\mu \gg m$,  the Compton wavelength of $\phi$ is the longest length scale of the problem.  Thus, it is valid to treat $\phi$ as a classical field driven by a localized source of $\chi$ particles.  In analogy with the analysis of \Sec{sec:point}, we can solve the equations of motion in search of radially symmetric, stable solutions which interpolate to the metastable vacuum.  As before we impose Dirichlet boundary conditions at large radii so that the field interpolates to the metastable vacuum.  We also fix Dirichlet boundary conditions at the origin.  This choice is physically motivated: due to gravitational redshift, the field is tethered to its metastable value at the horizon of the black hole.  That said, our results will be insensitive to the choice of boundary conditions at the origin.\footnote{This insensitivity provides some support for our prescription of neglecting changes to the effective potential in the near-horizon region.}

\begin{figure}[t]\hspace*{-.5cm}
\centerline{\includegraphics[width=0.6\columnwidth]{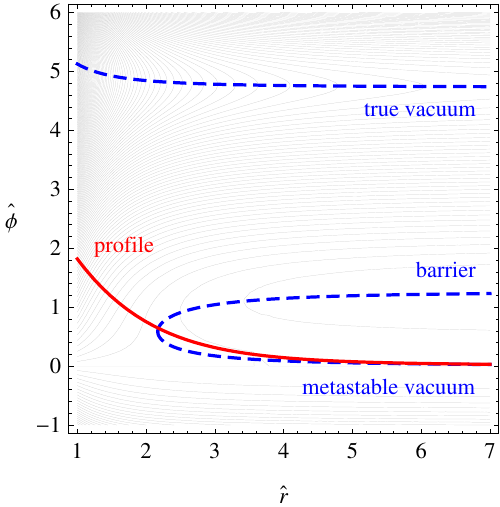}}
\caption{Contours of the full potential, $\hat V + \Delta \hat V_{\rm BH}$, as a function of the radius, $\hat r$, and the field value, $\hat \phi$, fixing $\hat a=2$ and $\hat b =1.3$.  The dashed blue lines label the positions of the metastable vacuum, stable vacuum, and potential barrier.  At finite radius the barrier merges with the metastable vacuum and disappears.  The solid red line depicts a stable $\hat \phi$ profile for this model. }\label{fig:Veff}
\end{figure}

We can understand the underlying physics by analyzing the shape of the potential near the black hole.  In dimensionless units, the full potential is $\hat V + \Delta \hat V_{\rm BH}$, where 
 \bea
 \Delta \hat V_{\rm BH} &=& \Delta V_{\rm BH} \lambda /m^4.
 \eea
\Fig{fig:Veff} depicts contours of  $\hat V + \Delta \hat V_{\rm BH}$ as a function of $\hat \phi$ and $\hat r$, fixing  $\hat a=2$ and $\hat b = 1.3$.  The finite density potential dominates at $\hat r = 0$ and is negligible as $\hat r \rightarrow \infty$.  The dashed blue curves label critical points at which $\partial (\hat V +\Delta \hat V_{\rm BH}) / \partial \hat \phi  =0$.  At large radii there are three such lines, indicating the positions of the metastable vacuum, stable vacuum, and potential barrier.  Near the black hole, two of these lines merge at a critical radius at which the metastable vacuum and the potential barrier are coincident.  At this inflection point the barrier disappears completely.  Naively, small field fluctuations near this critical radius are unstable to exponential growth, suggesting the onset of catalyzed vacuum decay.  This conclusion is incorrect, however---while the potential drives $\hat \phi$ towards the true vacuum, this is counteracted by the gradient energy cost of spatial variations of $\hat \phi$.  Hence, the stability of the vacuum depends on the relative strength of the gradient energy compared to the finite density potential.
For example, for the model parameters in in \Fig{fig:Veff}, the equations of motion support a stable solution for $\hat \phi$ which is depicted by the solid red line.  As $\hat b$ is increased, however, this stationary solution is slowly dragged upwards to larger field values.  Past a certain critical value of $\hat b$, the solution ``snaps'' and the equations of motion can no longer support a stable solution.  In this case vacuum decay is classically catalyzed by the black hole.

To determine the critical value of $\hat b$ let us study this system in various simplifying limits.  For example, consider the theory as $\hat a \gg \sqrt{3}$, corresponding to a weak potential barrier.  In empty space, the associated vacuum tunneling transition is mediated by a thick wall bubble.  In the large $\hat a$ limit, the dynamics are  independent of the quartic stabilization term in the potential.  Dropping the $\hat \phi^3$ term in \Eq{eq:eom}, it is clear that the equations of motion depend on the model parameters  in the specific combination $\hat a \hat b$.  Thus, at large $\hat a$, catalyzed vacuum decay will occur above a critical value 
\bea
\hat b &>& \frac{\textrm{constant}}{\hat a},
\label{eq:bcrit1}
\eea
with a positive proportionality constant which is difficult to compute analytically.
This result is physically reasonable: larger values of $\hat a$ imply a smaller potential barrier and larger values of $\hat b$ imply stronger finite density effects.

Alternatively, consider the opposite limit, $\hat a \rightarrow \sqrt{3}$, in which the metastable and stable vacua are approximately degenerate.  In empty space, the vacuum decay transition is mediated by a thin wall bubble.  We can crudely characterize the strength of catalysis by computing the effective potential for a collective coordinate  labeling the radius of a nucleated thin wall bubble.  In particular, consider a thin wall ansatz of the form
\bea
\hat\phi &=& \left\{ \begin{array}{ll}
0&, \quad \hat r > \hat R\\
\hat \phi_{\rm true} &, \quad \hat r < \hat R
\end{array}
\right.
\eea
where $\hat R = Rm$ is the bubble wall radius in dimensionless units  and  $\hat\phi_{\rm true} =2\sqrt{3}$ is the true vacuum in the thin wall limit.  Integrating the induced tadpole $-\hat b /\hat r^2$ over the bubble ansatz, we obtain the finite density contribution to the effective potential for the bubble wall,
\bea
\hat V_{\rm bubble} &=& V_{\rm bubble} \lambda/m^4 \nonumber \\
&=& 4\pi \hat R^2 \hat \sigma- \frac{4\pi \hat R^3 \hat \epsilon}{3}  -4\pi  \hat R \hat b  \hat \phi_{\rm true},
\label{eq:Vbubble}
\eea
where $\hat \sigma = \int_0^{\hat \phi_{\rm true}} d\hat \phi\; (2 \hat V)^{1/2}  = 2$ and $\hat \epsilon =  4\sqrt{3}(\hat a -\sqrt{3})$ are the surface tension and vacuum energy difference, respectively, in the thin wall limit.  The potential barrier in \Eq{eq:Vbubble} disappears when
\bea
\hat b > \frac{\hat \sigma^2}{\hat \epsilon \hat \phi_{\rm true}} = \frac{1}{6(\hat a-\sqrt{3})},
\label{eq:bcrit2}
\eea
which diverges in the thin wall limit.  Thus, very large $\hat b$ is needed to catalyze vacuum decay in this regime.  Physically, this is reasonable because a thin wall bubble nucleates with a radius parametrically greater than the microphysical scale of the $\phi$ potential.  Hence, to catalyze vacuum decay, the total flux of emitted quanta must be commensurately higher to induce a sizable finite density potential at such large radii.

\begin{figure}[t]\hspace*{-.5cm}
\centerline{\includegraphics[width=0.7\columnwidth]{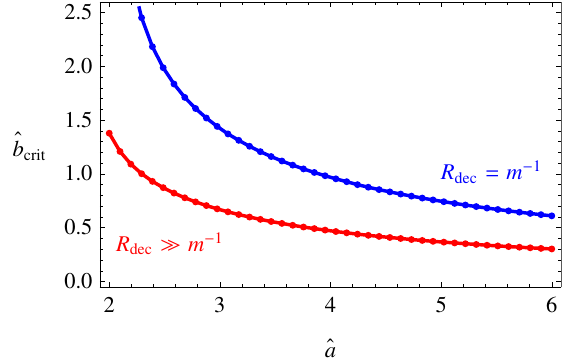}}
\caption{Critical values of model parameters, above which vacuum decay of $\phi$ is classically catalyzed by a black hole.  The red (blue) line corresponds to emitted $\chi$ particles which have a decay length much longer than (equal to) the Compton wavelength of $\phi$.  Large values of $\hat b > \hat b_{\rm crit}$ typically arise in theories with hierarchical masses.  The thick wall ($\hat a \gg \sqrt{3}$) and thin wall ($\hat a \rightarrow \sqrt{3}$) limits are consistent with the analytic predictions of \Eq{eq:bcrit1} and \Eq{eq:bcrit2}.
}\label{fig:bcrit}
\end{figure}

Our analytic results are consistent with a numerical analysis of the equations of motion.  We have scanned the parameter space, identifying all points that support stable, radially symmetric solutions.  The resulting phase diagram is presented in \Fig{fig:bcrit}.  The red line indicates $\hat b_{\rm crit}$, the critical value of $\hat b$ above which the vacuum is classically unstable.  As predicted, for $\hat a \gg \sqrt{3}$, the critical boundary asymptotes to the form in \Eq{eq:bcrit1};  for $\hat a \rightarrow \sqrt{3}$, the critical boundary diverges as suggested by \Eq{eq:bcrit2}.  

To compute the position of the red line in \Fig{fig:bcrit} we have assumed that $\chi$ is stable.  This result is robust, provided the decay length of the emitted $\chi$ particles, $R_{\rm dec} = \Gamma_{\rm dec}^{-1} \gamma$, is much greater than $m^{-1}$, the characteristic length scale of the $\phi$ potential.  This is often the case because the emitted quanta are relativistic and $\gamma \gg 1$.  However, for sufficiently heavy $\chi$ particles, the decay length may be of order or shorter than the Compton wavelength of $\phi$.  In this scenario, the profile of $\chi$ flux is modified by the replacement
\bea
 \frac{1}{r^2} &\rightarrow & \frac{e^{-r/R_{\rm dec}}}{r^2},
 \label{eq:cutreplace}
 \eea
  and the finite density potential is suppressed at large radii.  As a result, catalyzed vacuum decay is more difficult to achieve.  To illustrate the effect of $\chi$ decays, we have computed $\hat b_{\rm crit}$ assuming $R_{\rm dec} = m^{-1}$, which is depicted by the  blue line in \Fig{fig:bcrit}.  Interestingly, even if $\chi$ decays, black hole catalyzed vacuum decay can still occur.  That said, for $R_{\rm dec} \ll m^{-1}$, $\hat b_{\rm crit}$ diverges and catalysis shuts off.

Finally, let us comment on the effects of the $\phi$ quanta emitted by the black hole. Setting $\chi=\phi$ in the above analysis, we find that
\bea
\hat{b} &=& \frac{fa\sqrt{\lambda}}{m}~.
\eea
Since $\hat{b}>0$, $\phi$ is driven toward the true vacuum.  However, because $f \ll 1$ and $a\sim m$, we always find $\hat{b} \ll 1$. It is clear from \Fig{fig:bcrit} that we need $\hat{b} \sim {\cal O}(1)$ to catalyze vacuum decay, so $\phi$ quanta are negligible for catalyzed vacuum decay.

\subsubsection{Tachyon Instability}

\label{sec:tachyon}

Next, we consider a scenario where the leading instability of the  finite density potential is a tachyon.  This requires that $\mu^{2\,\prime}=0$ and $\mu^{2\, \prime\prime} <0$, so the tadpole contribution vanishes.  This occurs if the dynamics are invariant under a parity of the scalar field, $\phi \rightarrow -\phi$, which is automatic when $\phi$ is a component of a charged multiplet.  This parity constrains the equations of the motion to the form
\bea
 \hat\square \hat \phi + \left(1-  \frac{\hat c }{\hat  r^2}  \right) \hat \phi +  {\cal O}(\hat \phi^3) &=&0,
 \label{eq:eomtachyon}
\eea
where $\hat c >0$ controls the strength of the induced tachyon,
\bea
\hat c &=& -f \mu^{2\, \prime\prime}.
\label{eq:cdef}
\eea
As we will see, the potential beyond quadratic order will be unimportant for the coming discussion.  Naively, \Eq{eq:eomtachyon} suggests that small fluctuations of $\phi$ are unstable to a localized tachyon in a picture reminiscent of the localized tadpole in \Eq{eq:eom}.  To understand whether this instability catalyzes vacuum decay, let us compute the general solution of \Eq{eq:eomtachyon} in the  linearized limit.  We assume a radially symmetric ansatz,
\bea
\hat \phi(\hat t,\hat r) &=&\frac{ e^{-i \hat \omega \hat t}}{\hat r} U(\hat r),
\label{eq:ansatz}
\eea
where $\hat \omega$ is the frequency in dimensionless units.  The equation of motion can be massaged into the form of a Schrodinger equation,
\bea
\hat \omega^2 U(\hat r) &=& \left[ - \frac{\partial^2 }{\partial \hat r^2} + \left(1- \frac{ \hat c}{\hat r^2}\right) \right] U(\hat r).
\label{eq:schrod}
\eea
Applying the quantum mechanical analogy, we identify the quantity in square brackets as the ``Hamiltonian''.  If this Hamiltonian supports negative energy bound state solutions, then $\hat \omega^2 <0$ and the ``ground state'' of the system corresponds to a configuration with imaginary frequency.   This signals a true tachyonic instability in the theory.      
\Eq{eq:schrod} has a general analytic solution which is real and vanishing at infinity,
\bea
U(\hat r) &\propto & \sqrt{\varepsilon \hat r} K_{\beta}(\varepsilon \hat r),
\label{eq:gensol}
\eea
where $\beta =\sqrt{1/4-\hat c}$ and  $\varepsilon^2 = 1-\hat\omega^2 $.    Because $U(\hat r)$ is a bound state solution, $\varepsilon$ should be a quantized.  However, it appears as a continuous parameter labeling the eigenmodes of \Eq{eq:schrod}.  This is a sign of the underlying conformal symmetry of the potential, whereby space and time rescale uniformly.   With no dimensionful parameters to provide a gap for the discretuum, the spectrum of eigenmodes is continuous and unbounded from below.  However, in any realistic physical system, the potential is regulated at small radii by a physical short distance scale.   For the realistic system of a black hole, this regulator is the Schwarzschild radius.  More generally, fixing a boundary condition at some small radius will discretize $\varepsilon$, provided that $U(\hat r)$ has nodes.  This only happens if the solution is oscillatory, 
which is only possible if $\beta$ is imaginary, so
\bea 
\hat c &>& 1/4.
\label{eq:tachyoncond}
\eea
This condition is well known in the context of conformal quantum mechanical systems  \cite{QMpaper}.  
\Eq{eq:tachyoncond} is reasonable because the localized tachyon must overpower the gradient energy cost required to destabilize the field.  However,  this inequality is difficult to satisfy because according to \Eq{eq:fTexact}, $f$ is small, and 
$\mu^{2\, \prime\prime}$ is proportional to perturbative couplings.  Thus,  \Eq{eq:tachyoncond} is never satisfied and the theory does not support an unstable mode; vacuum decay is not catalyzed by black hole induced tachyons.

\begin{figure}[t]\hspace*{-.5cm}
\centerline{\includegraphics[width=0.7\columnwidth]{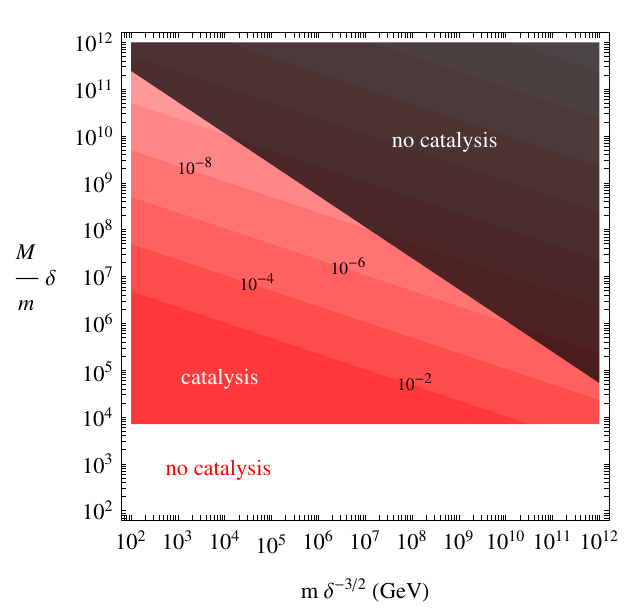}}
\caption{Phase diagram of the parameter space constrained by black hole catalyzed vacuum decay for the explicit model in \Eq{eq:Lexplicit}.  There is no catalysis in the white region because the finite density potential induced by emitted $\chi$ particles cannot overcome the potential barrier for $\phi$.  There is no catalysis in the black region because black holes that are sufficiently hot to produce $\chi$ particles decay too fast to influence the $\phi$ dynamics.   }\label{fig:pheno}
\end{figure}

\section{Phenomenological Implications}

\label{sec:pheno}




In this section we formulate the necessary conditions for black hole catalyzed vacuum decay.  For a scalar $\phi$ sitting at a metastable vacuum, decay is catalyzed by Hawking radiated $\chi$ particles if the following criteria are simultaneously satisfied:

\begin{itemize}[leftmargin=.5cm,rightmargin=.5cm] 

\item[] {\it i}) At least one black hole has decayed in our past light cone.  

\item[] {\it ii}) This black hole has a lifetime longer than the Compton wavelength of $\phi$.

\item[] {\it iii}) The emitted $\chi$ quanta have a decay length longer than the Compton wavelength of $\phi$.

\item[] {\it iv}) These particles couple to $\phi$ appropriately to drive $\phi$ over its potential barrier.

\end{itemize}
For our analysis, we simply assume condition {\it i}).  Crucially, the Hawking temperature of the decaying black hole scans up to the Planck scale, inevitably crossing the threshold for $\chi$ particle production, $T \gtrsim M$.  To influence the evolution of $\phi$, however, this black hole must survive for a period longer than the characteristic wavelength set by the $\phi$ potential.  This is condition {\it ii}), which is often satisfied because the black hole lifetime scales inversely with the strength of gravity.  Given that the black hole is sufficiently long-lived, a cloud of Hawking radiation forms around it.  For stable $\chi$ particles, this halo forms a $1/r^2$ profile, while for unstable $\chi$ particles the distribution dissipates with an extra factor of $e^{-r/R_{\rm dec}}$, where $R_{\rm dec}$ is the $\chi$ decay length.  Condition {\it iii}) guarantees that the phase space distribution of $\chi$ is non-zero at scales of order the Compton wavelength of $\phi$.  Otherwise, the finite density potential induced by the $\chi$ quanta will be too weak to affect the $\phi$ dynamics.  Since the emitted $\chi$ particles are relativistic, $R_{\rm dec}$ is enhanced by a substantial boost factor which makes this condition more easily satisfied.   Lastly, condition {\it iv}) says that the couplings of $\phi$ to $\chi$ must have the appropriate sign and magnitude to drive $\phi$ over its potential barrier.


\subsection{Example Model}

\label{sec:exmodel}
  
Safeguarding the vacuum from catalyzed decay implies new constraints on particle physics models.  
For concreteness, consider an explicit model defined by the scalar potential in \Eq{eq:Vexplicit}, together with the interaction term in \Eq{eq:Lint}.  Presented all together, the Lagrangian is:
\bea
{\cal L} &=& \frac{\partial \phi^2}{2} - \left( \frac{m^2 \phi^2}{2}  - \frac{a \phi^3}{3!} + \frac{\lambda\phi^4}{4!} \right) + \bar \chi i\sslash{\partial} \chi -\bar \chi (M-y \phi)\chi,
\label{eq:Lexplicit}
\eea
where $a > \sqrt{3\lambda }m $ so that the origin is metastable; the true vacuum is at positive values of $\phi$.  
If condition {\it i}) is true, then $\chi$ particles start to be emitted by the black hole as soon as $T\gtrsim M$.  However, hotter black holes decay faster, which is in tension with condition {\it ii}), the criterion that $\tau _{\rm BH}\gtrsim m^{-1}$ where $\tau_{\rm BH}$ is the black hole lifetime defined in \Eq{eq:BHdecay}.  Plugging $T \gtrsim M$ into condition {\it ii}), we find that

\bea
M &\lesssim& \left(\frac{10 \,m \, \mPl^2 }{\pi^2n_*}\right)^{1/3},
\label{eq:BHdeclimit}
\eea
which is a necessary condition for black hole catalyzed vacuum decay.  Here $n_*$ is the effective number of massless degrees of freedom emitted by the black hole, accounting for the differences in grey body factors between particles of different spin~\cite{Page:1976df}. Without knowledge of the full spectrum beyond the standard model, we cannot say for certain how large $n_*$ is. However, for $n_*\lesssim 10^3$ it makes little difference. Physically, \Eq{eq:BHdeclimit} is reasonable:  if $M$ is too large, then black holes that are hot enough to produce $\chi$ will decay too quickly to influence the evolution of $\phi$.  
Given the Lagrangian in \Eq{eq:Lexplicit}, $\chi$ is stable and condition {\it iii}) is thus satisfied.  We will return to the possibility of unstable $\chi$ shortly.

If $y$ is positive, then $\phi$ is driven to positive values in order to decrease the mass of the emitted $\chi$ particles. 
According to \Eq{eq:Meff}, the effective mass is $\mu^2[\phi] = (M-y\phi)^2$.  Plugging into
\Eq{eq:bdef}, we find that condition {\it iv}) is satisfied if
 \bea
y >0 &\textrm{  and  }& M >  \frac{m}{2f \delta} .
\label{eq:hierarchy}
\eea
Here we have defined a quantity $\delta = y \sqrt{\lambda}/\hat b_{\rm crit}$ which is order one or smaller, where $\hat b_{\rm crit}$ is indicated by the red line in \Fig{fig:bcrit}.   
Since $f \ll 1$, catalysis requires a hierarchy between $M$ and $m$.  Note that in this regime, \Eq{eq:denseenough} is safely satisfied when $T\gtrsim M$ at radii $r \sim m^{-1}$, so the mean field approximation for the $\chi$ phase space density is justified.

Interestingly, black hole catalyzed vacuum decay places a stringent constraint on large hierarchies among interacting states: a stable vacuum requires $M$ to sit outside a window bounded from above by \Eq{eq:BHdeclimit} and from below by \Eq{eq:hierarchy}.  In \Fig{fig:pheno}, the large shaded red region satisfies conditions {\it i}) - {\it iv}) and is subject to catalyzed vacuum decay.  The black region violates \Eq{eq:BHdeclimit} because black holes sufficiently hot to produce $\chi$ particles decay too fast to affect the $\phi$ field evolution.  The white region violates \Eq{eq:hierarchy} because the $\chi$ mass is too small to overcome the potential barrier. 

While $\chi$ is stable for the theory defined in \Eq{eq:Lexplicit}, it decays in many realistic models where condition ${\it iii})$ is not automatic.  For unstable $\chi$ particles, the phase space distribution of $\chi$ dissipates according to the replacement in \Eq{eq:cutreplace}.   The decay length of $\chi$ is $R_{\rm dec} = \Gamma_{\rm dec}^{-1} \gamma $, which can be significantly enhanced by the boost factor $\gamma \sim T / M$, as discussed in detail in \Sec{sec:tadpole}.  To see how $\chi$ decays weaken our limits,  we have included contours of red bands in \Fig{fig:pheno} indicating critical values of $\Gamma_{\rm dec} / M\delta $, above which $R_{\rm dec} \leq m^{-1}$, so the decay length is shorter than the Compton wavelength of $\phi$ and catalysis shuts off.  For prompt decays, the instability region is smaller, but still viable, especially at smaller values of $m$.  On the other hand, as the decays become longer-lived---say if they are mediated through higher dimension operators---then a greater portion of the parameter space falls victim to catalyzed vacuum decay.

\subsection{Standard Model}

\label{sec:SM}

At tree-level, the SM Higgs potential supports a unique minimum: the electroweak symmetry breaking vacuum.  However, as is well-known the vacuum structure is enriched at loop-level \cite{Arnold:1989cb,Arnold:1991cv}.  Taken at face value, the observed Higgs mass \cite{Aad:2012tfa,Chatrchyan:2012ufa} implies that the Higgs quartic runs negative at an intermediate scale of order  $10^{10} \textrm{ GeV}$ \cite{Ellis:2009tp,EliasMiro:2011aa}.    Strictly speaking, the Higgs potential is unbounded from below at this scale and the quantum theory does not have a ground state.  Of course, in any well-defined ultraviolet completion, this unbounded field direction is lifted by higher dimension operators.  In fact, negative quartic couplings are a natural byproduct of integrating out additional scalar fields.  

Naively, this instability is quite severe.  The mass is tachyonic and the quartic coupling is negative.  However, \cite{Lee:1985uv} famously showed that even in the absence of a potential barrier, fluctuations of the Higgs are classically stable, on account of the substantial gradient energy cost of nucleating a bubble of true vacuum.  Instead, the Higgs must quantum mechanically tunnel from the electroweak vacuum.  The resulting decay rate is exponentially suppressed and thus the lifetime of the electroweak vacuum is longer than the age of the universe \cite{Ellis:2009tp,EliasMiro:2011aa}.  Hence, the SM Higgs lies in an apparent region of metastability, but in a way that is consistent with observation.  Note that this conclusion is subject to important experimental uncertainties on the top quark mass and strong gauge coupling.

If the electroweak symmetry breaking minimum is indeed metastable, then it is reasonable to ask whether vacuum decay can be catalyzed by a black hole.  
Since $\phi$ is a component of an electroweak doublet, the scalar potential preserves a $\phi \rightarrow -\phi$ parity.  Hence, the finite density potential is even in $\phi$, so $\mu^{2\, \prime} =0$ and the leading contribution is quadratic.  In the SM, the dominant contributions to the finite density potential come from the top quark, the electroweak bosons, and the Higgs itself.  Since these particles acquire mass entirely from electroweak symmetry breaking, larger values of the Higgs will increase their effective mass.  Thus, $\mu^{2 \, \prime\prime }>0$ and the induced potential tends to push $\phi$ towards the origin of field space.  While this phenomenon may have interesting implications for restoration of electroweak symmetry, {\it e.g.}~for the purposes of electroweak baryogenesis  \cite{Nagatani:1998gv}, it does not drive the field in the direction of the quartic instability.  Black holes do not catalyze decay of the electroweak vacuum in the SM.

\subsection{Beyond the Standard Model}

\label{sec:BSM}

In \Sec{sec:exmodel} we derived new stability limits on a simple scalar model.  We then argued in \Sec{sec:SM} that catalyzed decay does not constrain the SM.  What about constraints on motivated extensions of the SM?  Trivially, our limits apply to the SM augmented by a singlet scalar $\phi$ with the potential in \Eq{eq:Vexplicit}.  However, similar constraints also apply to metastable vacua in the singlet-extended SM \cite{Gonderinger:2009jp,Gonderinger:2012rd} and the next-to-minimal supersymmetric standard model (NMSSM) \cite{Kanehata:2011ei,Kobayashi:2012xv,Agashe:2012zq}.  Depending on the precise couplings between the Higgs and the singlet, a black hole can drive the Higgs from the electroweak symmetry breaking vacuum.   A proper treatment of this phenomenon will likely require a multi-field analysis beyond the scope of the present work.

We have seen that black hole catalyzed vacuum decay places stringent limits on light scalars coupled to heavy particles.  As emphasized in \cite{Green:2006nv} this scenario automatically arises when $\phi$ is the pseudo-Goldstone boson of a spontaneously broken symmetry.  Here $\chi$ will be parametrically heavier than $\phi$, provided its mass is unprotected by the preserved symmetry group.   For example, we could identify $\phi$ with the radion field parameterizing a de-compactification transition to a higher dimensional vacuum, and $\chi$ with the associated Kaluza-Klein particles.  

Such mass hierarchies also arise in models of metastable supersymmetry (SUSY) breaking.  As discussed in \cite{Ray:2006wk,Komargodski:2009jf}, at tree-level, the SUSY breaking modulus is generically a flat direction in field space.  It is then natural to identify $\phi$ with the SUSY breaking modulus and $\chi$ with heavier messenger states.   Without loss of generality, we can shift $\phi $ so that its vacuum expectation value is at the origin; in analogy with \Eq{eq:Lint}, its couplings become
\bea
W_{\rm int} &=& (M - y \phi) \bar \chi \chi.
\label{eq:Wint}
\eea
An evaporating black hole will emit a flux of $\chi$ messenger particles which induces a finite density potential for the SUSY breaking modulus $\phi$.   If conditions {\it i}) - {\it iv}) are satisfied, then the metastable SUSY breaking vacuum is unstable to black hole catalyzed vacuum decay.  It remains to be seen whether these conditions are consistent with detailed model building constraints such as $R$-symmetry breaking.  We leave a more detailed analysis for future work.


\label{sec:limits}

\section{Conclusions}
\label{sec:conclusions}

We have presented a systematic analysis of vacuum decay induced by ambient matter.  For perturbative theories, it is difficult for a single particle $\chi$ to drive $\phi$ over its potential barrier.  As shown in \Fig{fig:ycrit}, catalysis only occurs if these states are strongly coupled or if the barrier is very weak.  On the other hand, catalysis is much easier in the presence of many $\chi$ quanta, for example as would result from the Hawking radiation of a black hole.
In \Eq{eq:Veff}, we have presented the finite density potential induced for $\phi$ by the $\chi$ quanta emitted by a black hole.  Because we are interested in distances far from the event horizon, subtleties about the information paradox are irrelevant to our analysis---the black hole is simply a source of high energy particles, much like star.  If the basin of attraction of the finite density potential is sufficiently misaligned from the metastable vacuum, then $\phi$ can be driven over its potential barrier, catalyzing decay.   The critical values of couplings at which catalysis occurs are shown in \Fig{fig:bcrit}.  Finally, we have summarized the necessary conditions for black hole catalyzed vacuum decay in the beginning of \Sec{sec:pheno}.  Demanding vacuum stability implies new constraints on theories with hierarchical spectra, {\it e.g.}~as shown in \Fig{fig:pheno} for the model defined \Eq{eq:Lexplicit}.  

This work leaves many avenues for future work.   As noted in \Sec{sec:BSM}, first and foremost is a comprehensive analysis of new limits on beyond the SM theories, {\it e.g.}~singlet extensions of the SM, the NMSSM, the radion, and metastable SUSY breaking models.   Stability bounds will have the most significance for theories with hierarchical spectra.   Moreover, our findings could have more general implications for the landscape: vacua with larger mass hierarchies are in greater danger of catalyzed vacuum decay.  

Throughout, our discussion has assumed the decay of at least one black hole in our past light cone.  As is well-known, primordial black holes can be produced by over-densities of curvature perturbations after inflation, as well as during first order phase transitions.  It would be interesting to understand the likelihood that exactly zero primordial black holes were produced in our past light cone.  

Lastly, it may be fruitful to consider other applications of the finite density potential induced by a black hole.  In principle, this potential can drive the $\phi$ potential into a more symmetric phase.  Effectively, this forms a domain wall surrounding the black hole.  This field configuration could accommodate a novel mechanism for baryogenesis, {\it e.g.}~if electroweak symmetry or a grand unified symmetry is restored.

\begin{center}{\bf \large Acknowledgements}\end{center}
 
We would like to thank Sean Carroll, I-Sheng Yang, and Mark Wise for helpful comments, and we are especially grateful to Paul Steinhardt for collaboration in the early stages of this work.   This research is supported by the DOE under contract \#DE-FG02- 92ER40701 and the Gordon and Betty Moore Foundation through Grant \#776 to the Caltech Moore Center for Theoretical Cosmology and Physics.  C.C~is supported by a DOE Early Career Award \#DE-SC0010255 and S.L.~is supported by a John A. McCone Postdoctoral Fellowship.  C.C.~would also like to thank the Aspen Center for Physics and the Kavli Institute for Theoretical Physics in Santa Barbara, where part of this work was completed.


\begin{thebibliography}{}


\bibitem{Khlebnikov:1986ky} 
  S.~Y.~.Khlebnikov and M.~E.~Shaposhnikov,
ÊÊPhys.\ Lett.\ B {\bf 180}, 93 (1986).
ÊÊ

\bibitem{Dimopoulos:1990at} 
  S.~Dimopoulos, B.~W.~Lynn, S.~B.~Selipsky and N.~Tetradis,
ÊÊPhys.\ Lett.\ B {\bf 253}, 237 (1991).
ÊÊ

\bibitem{Anderson:1990kb} 
  G.~W.~Anderson, L.~J.~Hall and S.~D.~H.~Hsu,
ÊÊPhys.\ Lett.\ B {\bf 249}, 505 (1990).
ÊÊ
  
\bibitem{Green:2006nv} 
  D.~R.~Green, E.~Silverstein and D.~Starr,
  Phys.\ Rev.\ D {\bf 74}, 024004 (2006)
  [hep-th/0605047].


\bibitem{Affleck:1979px} 
  I.~K.~Affleck and F.~De Luccia,
ÊÊPhys.\ Rev.\ D {\bf 20}, 3168 (1979).
ÊÊ

\bibitem{Voloshin:1993dk} 
  M.~B.~Voloshin,
ÊÊPhys.\ Rev.\ D {\bf 49}, 2014 (1994).
ÊÊ


\bibitem{Berezin:1990qs} 
  V.~A.~Berezin, V.~A.~Kuzmin and I.~I.~Tkachev,
ÊÊPhys.\ Rev.\ D {\bf 43}, 3112 (1991).
ÊÊ

\bibitem{Arnold:1989cq} 
  P.~B.~Arnold,
ÊÊNucl.\ Phys.\ B {\bf 346}, 160 (1990).
ÊÊ

\bibitem{Hawking:1974rv} 
  S.~W.~Hawking,
ÊÊNature {\bf 248}, 30 (1974).
ÊÊ

\bibitem{Hawking:1974sw} 
  S.~W.~Hawking,
ÊÊCommun.\ Math.\ Phys.\  {\bf 43}, 199 (1975)
ÊÊ[Erratum-ibid.\  {\bf 46}, 206 (1976)].
ÊÊ
  
\bibitem{Unruh:1976fm} 
  W.~G.~Unruh,
ÊÊPhys.\ Rev.\ D {\bf 14}, 3251 (1976).
ÊÊ

\bibitem{Page:1976df} 
  D.~N.~Page,
  Phys.\ Rev.\ D {\bf 13}, 198 (1976).
  
\bibitem{MacGibbon:2007yq} 
  J.~H.~MacGibbon, B.~J.~Carr and D.~N.~Page,
  Phys.\ Rev.\ D {\bf 78}, 064043 (2008)
  [arXiv:0709.2380 [astro-ph]].
  
\bibitem{Harris:2003eg} 
  C.~M.~Harris and P.~Kanti,
  JHEP {\bf 0310}, 014 (2003)
  [hep-ph/0309054].
  
\bibitem{Cvetic:1997ap} 
  M.~Cvetic and F.~Larsen,
  Phys.\ Rev.\ D {\bf 57}, 6297 (1998)
  [hep-th/9712118].



\bibitem{QMpaper}
  S.~A.~Coon, B.~R.~Holstein,
  [quant-ph/0202091].



\bibitem{Coleman:1977py} 
  S.~R.~Coleman,
ÊÊPhys.\ Rev.\ D {\bf 15}, 2929 (1977)
ÊÊ[Erratum-ibid.\ D {\bf 16}, 1248 (1977)].
ÊÊ

\bibitem{Callan:1977pt} 
  C.~G.~Callan, Jr. and S.~R.~Coleman,
ÊÊPhys.\ Rev.\ D {\bf 16}, 1762 (1977).
ÊÊ

\bibitem{Linde:1980tt} 
  A.~D.~Linde,
ÊÊPhys.\ Lett.\ B {\bf 100}, 37 (1981).
ÊÊ

\bibitem{Cheung:2012nb} 
  C.~Cheung, M.~Papucci and K.~M.~Zurek,
ÊÊJHEP {\bf 1207}, 105 (2012)
ÊÊ[arXiv:1203.5106 [hep-ph]].
ÊÊ


\bibitem{Arnold:1989cb} 
  P.~B.~Arnold,
ÊÊPhys.\ Rev.\ D {\bf 40}, 613 (1989).
ÊÊ

\bibitem{Arnold:1991cv} 
  P.~B.~Arnold and S.~Vokos,
ÊÊPhys.\ Rev.\ D {\bf 44}, 3620 (1991).
ÊÊ

\bibitem{Aad:2012tfa} 
  G.~Aad {\it et al.}  [ATLAS Collaboration],
  Phys.\ Lett.\ B {\bf 716}, 1 (2012)
  [arXiv:1207.7214 [hep-ex]].
  
\bibitem{Chatrchyan:2012ufa} 
  S.~Chatrchyan {\it et al.}  [CMS Collaboration],
  Phys.\ Lett.\ B {\bf 716}, 30 (2012)
  [arXiv:1207.7235 [hep-ex]].

\bibitem{Ellis:2009tp} 
  J.~Ellis, J.~R.~Espinosa, G.~F.~Giudice, A.~Hoecker and A.~Riotto,
ÊÊPhys.\ Lett.\ B {\bf 679}, 369 (2009)
ÊÊ[arXiv:0906.0954 [hep-ph]].
ÊÊ

\bibitem{EliasMiro:2011aa} 
  J.~Elias-Miro, J.~R.~Espinosa, G.~F.~Giudice, G.~Isidori, A.~Riotto and A.~Strumia,
ÊÊPhys.\ Lett.\ B {\bf 709}, 222 (2012)
ÊÊ[arXiv:1112.3022 [hep-ph]].
ÊÊ

\bibitem{Lee:1985uv} 
  K.~-M.~Lee and E.~J.~Weinberg,
  Nucl.\ Phys.\ B {\bf 267}, 181 (1986).

\bibitem{Nagatani:1998gv} 
  Y.~Nagatani,
ÊÊPhys.\ Rev.\ D {\bf 59}, 041301 (1999)
ÊÊ[hep-ph/9811485].
ÊÊ

\bibitem{Intriligator:2006dd} 
  K.~A.~Intriligator, N.~Seiberg and D.~Shih,
ÊÊJHEP {\bf 0604}, 021 (2006)
ÊÊ[hep-th/0602239].
ÊÊ

\bibitem{Intriligator:2007py} 
  K.~A.~Intriligator, N.~Seiberg and D.~Shih,
ÊÊJHEP {\bf 0707}, 017 (2007)
ÊÊ[hep-th/0703281].
ÊÊ
  
\bibitem{Gonderinger:2009jp} 
  M.~Gonderinger, Y.~Li, H.~Patel and M.~J.~Ramsey-Musolf,
ÊÊJHEP {\bf 1001}, 053 (2010)
ÊÊ[arXiv:0910.3167 [hep-ph]].
ÊÊ

\bibitem{Gonderinger:2012rd} 
  M.~Gonderinger, H.~Lim and M.~J.~Ramsey-Musolf,
ÊÊPhys.\ Rev.\ D {\bf 86}, 043511 (2012)
ÊÊ[arXiv:1202.1316 [hep-ph]].
ÊÊ

\bibitem{Kanehata:2011ei} 
  Y.~Kanehata, T.~Kobayashi, Y.~Konishi, O.~Seto and T.~Shimomura,
ÊÊProg.\ Theor.\ Phys.\  {\bf 126}, 1051 (2011)
ÊÊ[arXiv:1103.5109 [hep-ph]].
ÊÊ

\bibitem{Kobayashi:2012xv} 
  T.~Kobayashi, T.~Shimomura and T.~Takahashi,
ÊÊPhys.\ Rev.\ D {\bf 86}, 015029 (2012)
ÊÊ[arXiv:1203.4328 [hep-ph]].
ÊÊ

\bibitem{Agashe:2012zq} 
  K.~Agashe, Y.~Cui and R.~Franceschini,
ÊÊJHEP {\bf 1302}, 031 (2013)
ÊÊ[arXiv:1209.2115 [hep-ph]].
ÊÊ

\bibitem{Ray:2006wk} 
  S.~Ray,
ÊÊPhys.\ Lett.\ B {\bf 642}, 137 (2006)
ÊÊ[hep-th/0607172].
ÊÊ

\bibitem{Komargodski:2009jf} 
  Z.~Komargodski and D.~Shih,
ÊÊJHEP {\bf 0904}, 093 (2009)
ÊÊ[arXiv:0902.0030 [hep-th]].
ÊÊ
  
  \end{thebibliography}
\end{document}